\documentclass[12pt,A4,final]{iopart}
 \pdfoutput=1
\usepackage{graphicx}
\usepackage{amssymb}
\usepackage{color}
\usepackage{cite}   
\DeclareGraphicsExtensions{.pdf,.png,.jpg}

\usepackage{multirow}
\usepackage[%
  colorlinks=true,
  urlcolor=blue,
  linkcolor=blue,
  citecolor=blue,
  breaklinks=true
]{hyperref}

\begin{document}

\title{Jump Events in a 3D Edwards-Anderson Spin Glass}

\author{Daniel A. M\'{a}rtin$^\dag$, Jos\'e Luis Iguain$^\ddag$}
\ead{$\dag$danielalejandromartin@gmail.com, $\ddag$iguain@mdp.edu.ar}
\address{Instituto de Investigaciones F\'{\i}sicas de Mar del Plata (IFIMAR)
\\
CONICET and Universidad Nacional de Mar del Plata,\\
De\'an Funes 3350, 7600 Mar del Plata, Argentina}

\begin{abstract}
The statistical properties of infrequent particle displacements, greater 
than a certain distance, is known as jump dynamics in the context of 
structural glass formers. We generalize the concept of jump to the case of a 
spin glass, by dividing the system in small boxes, and considering infrequent 
cooperative spin flips in each box. Jumps defined this way share 
	similarities with jumps in structural glasses.
We perform numerical simulations for the 3D 
Edwards-Anderson model, and study how the properties of these jumps depend 
on the waiting time after a quench. Similar to the results for structural 
glasses, we find that while jump frequency depends strongly on time, jump 
duration and jump length are roughly stationary. 
At odds with some results reported on studies of structural glass formers,
at long enough times, the rest time between jumps varies as the
inverse of jump frequency. We give a possible explanation for this 
discrepancy. {We also find that our results are
	qualitatively reproduced by a fully-connected trap model.}
	\\
	\vspace{1pc}
\end{abstract}
\maketitle

\section{Introduction}

 Glass formers and spin glasses have been the subject of intense theoretical, 
experimental and numerical research in the last decades (see for instance, 
\cite{ReviewVidrios,SGPdestrians,SGRedux} and references therein).  
These systems present similar phenomenology (see for instance \cite{SGyG}), 
analogous techniques have been developed in both \cite{Iguain2008,SGyG2}, and 
their relationship has been proved \cite{KirkpatrickVidrioIgualSpin,BouchardVidrioigualSpin,ParisiVidrioigualSpin}.
Thus, in spite of the details about composition, structure or kind of degrees 
of freedom, it is common to use the word \emph{glassy} to refer to any 
material with a dynamics qualitatively similar to that of a 
glass~\cite{glassy_proc}.

The study of glassy systems poses a series of issues, related to the 
impossibility to reach equilibrium below the glass transition 
temperature $T_g$. 
In this case, the system evolves for ever. However, the time dependence of 
one-time observables, like density, energy, pressure or magnetization (in
a magnetic system) is quite weak and hard to appreciate. 
\emph{Aging} effects are more clearly observed in the behavior of two-time 
observables, like the autocorrelation
$C(t_2,t_1)$ and response $R(t_2, t_1)$ functions, which do not depend only 
on the time difference $t_2-t_1$ but on $t_1$ and $t_2$ (see, 
for example, \cite{refId0, 0305-4470-30-22-004, ANDREJEW1996117}). 

{
	Computer simulation methods have been of great help for the study of
	glasses. However, this approach is limited to rather small-scale
	problems, in comparison with characteristic times and lengths of
	real systems, which may diverge below $T_g$.
Some numerical techniques designed to avoid the effect of boundary 
conditions~\cite{TBCs}, or to overcome the slow dynamics~\cite{LulliParisi}, 
have been proposed. Also, special strategies that take profit of
conventional processors~\cite{MetodoInteresanteVMM} or
specific hardware~\cite{JANUS1,JANUS2} has been reported. 
Thanks to these advances, simulation times can be improved 
noticeably~\cite{JanusSecond}. Nevertheless, the asymptotic regimes in the 
glassy phase are still far from being reached.}

An alternative to sophisticated algorithms or expensive 
hardware is to investigate  
\emph{aging-to-equilibrium dynamics}~\cite{Aging2Eq,ElMasClaro}, 
{i. e., the out-of-equilibrium relaxation at a temperature above $T_g$.
When} a system is quenched  from an equilibrium state 
at a temperature $T_0$ to a lower temperature $T_F>T_g$, the transition to  
equilibrium takes place on a time scale related to the equilibrium 
relaxation time $\tau(T_F)$, { which diverges at $T_g$}.
{Thus, close to $T_g$ the relaxation is not trivial,
and  some aspects of glassy dynamics should  be revealed} for $t\ll \tau(T_F)$.
In \cite{ElMasClaro}, two-time correlation functions were studied in both  
aging-to-equilibrium and  aging regimes  for a  simple glass former. 
In  \cite{Aging2Eq}, the authors analyzed the aging-to-equilibrium dynamics 
for the strong glass former SiO$_2$.

Most of the studies about structural glass former systems focus on  \emph{macroscopic} or
global observables. Nevertheless, in recent years \emph{microscopic} 
actions have been 
analyzed \cite{Microscopicsio2,WarrenYRottler1,PolyWyR, 
Vollmayr2015,RottlerSolo} in  structural glass formers.
{It has been experimentally observed that, close to
a glass transition,  
particles spend long  times moving around a  position, until they jump 
and start moving around another position (see, for 
example~\cite{ExperimColloid}, for colloids).  
This has motivated a lot of work on \emph{particle jumps} 
}(see \cite{JumpsInGlasses} and references therein).
    Roughly speaking,
  jump events are particle movements greater than 
a certain threshold. Different definitions have been proposed, 
but the statistical properties of jumps do not qualitatively depend on
these details.
Jumps involve a small number of particles, of the order of ten. 
{They can be related to high decorrelation 
regions~\cite{Hierarchy}, and are expected to accelerate the dynamics. Jump length is smaller than 
particle size~\cite{ExperimColloid,JumpsInGlasses}, and it 
slightly decreases with cooling.} 

The motivations for studying jumps are manifold.
Jump events are not only closely related  
to macroscopic evolution observables 
like diffusivity \cite{DifusionCajaCiamarra}
but constitute an important 
ingredient in the relaxation of glass formers.
They provide a bridge between structure and dynamics~\cite{Cage}, 
by giving a more
quantitative idea of the \emph{cage} effect. 
This is the case 
in~\cite{CageCiamarra}, where, for a 2D model at low temperature,
non jumping particles were more likely found in highly ordered environments. 
 The  relationship among jumps and dynamical heterogeneities has also been established~\cite{JumpsInGlasses,Hierarchy},
 in terms of a facilitation mechanism~\cite{Facilitation}.
 The statistical properties of jumps can be related to different competing theories~\cite{JumpsInGlasses}.
Also, aging in glasses was analyzed in terms of jumps by monitoring the heat 
transfer between system and thermal bath over a short time~\cite{Sibani2005}.

In  \cite{Microscopicsio2}, the microscopic dynamics of SiO$_2$
were analyzed  and related to the macroscopic dynamics.
Jump statistics where studied. The authors found that the number of jumps 
decreases strongly with $t$, the time elapsed since the quench, reaching  
equilibrium at times compatible with $\tau(T_f)$.
 Other properties of jumps, as average length, time duration, and 
 surprisingly rest time between consecutive jumps did not depend on $t$. 
 In \cite{WarrenYRottler1}, a distinction among fist hop time (the time until
 the first jump) and persistence. They find that first-hop time 
 depends on the waiting time, while persistence is independent of system age.
In contrast, in \cite{Siboni} the authors studied the distribution functions of 
the first-passage time, and the persistence time, and found that both quantities evolve with time.

In this work, we  study some microscopic aspects of the 
aging-to-equilibrium and aging dynamics of the well known 3D Edwards-Anderson (EA) 
spin-glass model \cite{EA1975,EASimul1976ching,EASimul1BINDER1976}. This model has been extensively studied in a macroscopic way, {but also} microscopic 
results  have been reported \cite{LeticiaCajas, WindowManssen}, including the existence of a 
backbone \cite{BackBoneEA}. 
Here we are not interested in understanding the long-time low-temperature regime
of a large-size system. We use this model as an example to study the dynamics of jumps in
a spin glass.
We start by generalizing the concept of jump, introduced originally for 
structural glass formers, to the case of a spin glass. We find that our definition of jumps shares some similarities with the one 
in structural glasses, particularly, it includes displacements that contribute to the relaxation
of the system.
We wish to analyze whether the statistical properties of jump 
events reported in~\cite{Microscopicsio2} can be reproduced in a glassy system composed of spins, instead of 
moving particles.

For the EA model, we find that microscopic and macroscopic relaxation times behave 
similarly. We also find that some variables depend strongly on time
while others are nearly stationary.
Thus, most of the results in \cite{Microscopicsio2} for a structural glass former
hold for this spin glass model. For instance,
jump length and jump duration are roughly stationary, while the number of 
jumps 
depends strongly on $t$.  On the other hand, we find that {rest time} between jumps, which 
is independent of time in \cite{Microscopicsio2}, has an inverse relation to jump frequency 
for the EA model. We draw a plausible explanation for this discrepancy. 
Finally, we also study jumps in a infinite dimensional trap model~\cite{Trap}, a simplified representation of glassy systems which neglects correlations 
among consecutive movements. Regarding the statistical properties of
individual jumps, we find a strong similarity between
this simple model and the 3D-EA model.

The paper is organized as follows. In Sec.~\ref{SeccionDescripcion}, we 
describe the model and observables. In Sec.~\ref{ResMacro}, we define 
an equilibration time based on macroscopic observables. Main results, i.e.,
evolution of microscopic observables, are presented in Sec.~\ref{ResMicro}. 
In { Sec.~\ref{Discus}  
we discuss our findings and future directions.}
In Sec.~\ref{Conclus}, draw our conclusions. 
In Appendix we explain  the procedure to estimate the equilibration time, and show the consistency
with the equilibration time for energy.

\section{Model and observables}
\label{SeccionDescripcion}

We study the Gaussian EA model in a cube with $L$ spins by side, under periodic boundary conditions, defined by the Hamiltonian:
\begin{equation}
 H=-\sum_{\langle i,j\rangle} J_{ij} S_i S_j, \nonumber
\end{equation}
where the indexes $i$, $j$ run from $1$ to $L^3$. Spin variables take values 
$S_i=\pm 1$ and the pairs $\langle i,j\rangle$ identify nearest-neighbors. 
The couplings are taken randomly with a Gaussian distribution 
with zero mean and unit variance.

Dynamics are simulated with Metropolis algorithm. That is, 
at each Monte Carlo (MC) step, $N=L^3$ spin flip trials are performed. They 
are accepted or rejected according to their Boltzmann weights.

For this system, $T_g\simeq 0.95$ \cite{Tgnew}. To study aging-to-equilibrium, and also aging-to-non-equilibrium, we equilibrate the 
system at temperature $T_0=3$ (we have also performed runs for $T_0=\infty$) for $10^5$ MC steps. 
We define  $t=0$ as the time at which the system  is quenched to the final temperature $T_F=1.5,1.2,1.0$ and also $0.9$
which is below $T_g$.  We have run 1000 samples for each temperature, 
for $10^6$ MC steps, after the quench. 
{We have also run at least 60 samples for $10^8$ MC steps. We will focus on the range $10^3<t<10^6$ where most interesting phenomena takes place}.

In next paragraphs, we define several microscopic observables,
by dividing the full system into cubic boxes  of side $l_b$, 
which contain $N_b=l_b^3$ spins each.  
Unless explicitly stated, we have worked with $L=16$ and $l_b=4$.
Then, each box $C_i$ is labeled with index $i$ running from $1$ to $64$.
We record the configuration every $\delta t= 5$ MC steps 
which we will call a unit time. 

We define the overlap between consecutive records of box $i$ as
\begin{equation}
	O_{i}(t)=\frac{1}{N_b}\sum_{j\in C_{i}} S_j(t) S_j(t-\delta t)\,\,\,\mbox{,}
\end{equation}
 which we compute for $t=n\delta t$, with $n$ a positive integer.
 The magnetization $M_i$ and the energy $E_i$ of  every box are defined as

\begin{eqnarray}
	M_{i}(t)&=&\sum_{j\in C_{i}} S_j(t),\;\;\;\;\;\mbox{and}\\ 
	E_{i}(t)&=& {1\over 2}\sum_{j\in C_{i},k} J_{jk} S_j(t) S_k(t), 
\end{eqnarray}

where  $k$ is a nearest neighbor of $j$.

\begin{figure}[!ht]
 \centering 
   \includegraphics[scale=1.1, clip]{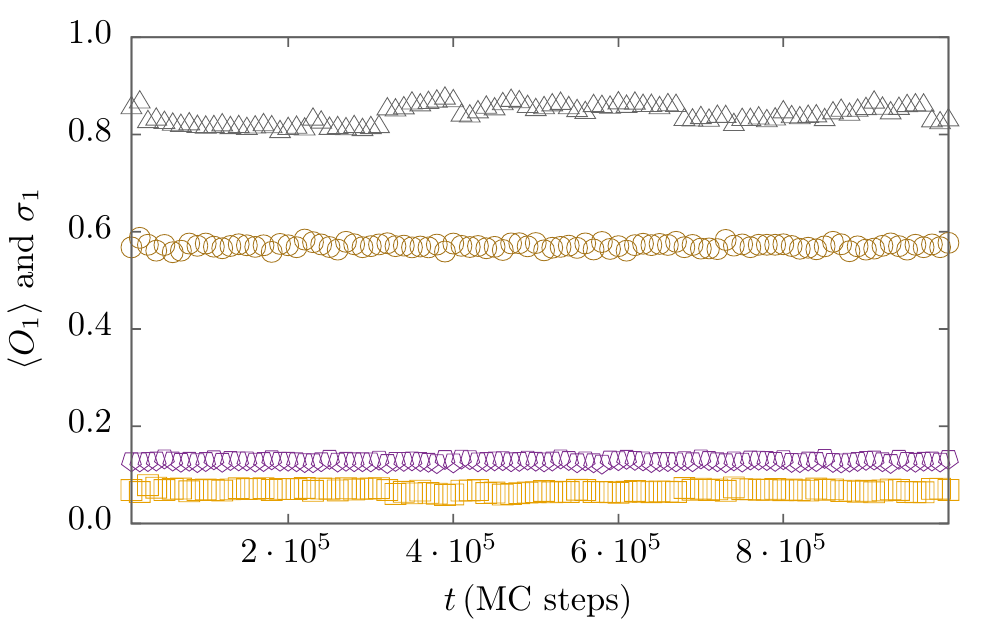}  
\caption{(Color online) $\sigma_1$ and $\langle O_1\rangle$ of a single box, 
 as a function of $t$,  for different temperatures. 
Black triangles: $\langle O_1\rangle$
for $T_F=0.9$. Grey circles: $\langle O_1\rangle$
for $T_F=1.5$. Orange squares: $\sigma_1$ for $T_F=0.9$. Violet pentagons:  $\sigma_1$ for $T_F=1.5$. }
 \label{Sigma}
 \end{figure}

When the simulation ends, we calculate $\langle O_{i}\rangle$ and $\langle O_{i}^2\rangle$, for all boxes averaged in the time window
$5. 10^5$ MC $<t<10^6$ MC.
With this, we calculate $\sigma_{i}=\sqrt{\langle O_{i}^2\rangle-\langle O_{i}\rangle^2}$.
In figure \ref{Sigma} we show the values of $O_{i}$ and $\sigma_{i}$ averaged over small time 
windows of $10 000$ MC steps, for a single box
($i=1$). We see that
both quantities are nearly constant even for the lowest temperature 
$T_F=0.9<T_g$. This makes them reasonably robust parameters to study jumps.

Greater changes in the configuration of a single box are related to smaller values of the overlap between two consecutive 
configurations. We will say that box  $i$ is jumping at time  $t$ if

\begin{eqnarray}
O_{i}(t) < \langle O_i\rangle -  \gamma \,\sigma_{i}.
\label{jumpdetect}
\end{eqnarray}
  
We have taken $\gamma=3$ in most of the cases, although we have also studied 
{{ $\gamma=2$ }} and $\gamma=5$. 
The jump starts at $t_i$, where $t_i$ is the greatest value that verifies
both $t_i<t$ and that Eq. (\ref{jumpdetect}) does not hold for $t_i$ (i.e. $O_{i}(t_i)\geq  \langle O_i\rangle -  \gamma \,\sigma_{i} $).
Similarly, the jump ends at $t_f>t$, if $t_f +\delta t $ is the smallest value
that does not verify Eq. (\ref{jumpdetect}).

We define the \emph{jump frequency} $\nu$, as the number of jumps per box 
per unit time. We measure it as a function of $t$.
For each jump, we define a \emph{jump duration} as $d=t_f-t_i$.   We define the 
\emph{rest time} $r$, as the time the particle waits until next jump. That is,
the difference between $t_i$ for the next jump and $t_f$
for the current jump. 
We define the \emph{jump size} $l$ using overlap values: we calculate the 
sum of the overlap changes over the times belonging to the jump,
i.e. $l=\sum_{t_i<t\leq t_f} N_b-O_{i}(t)$.

We also compute some macroscopic one-time quantities as the total energy, 
magnetization, the number of spins that flip, and the global two-time 
correlation $C(t+\Delta t,t)$:  

\begin{equation}
	C(t+\Delta t,t)={1\over N}\sum_{i=1}^{N} S_i(t) S_i(t+ \Delta t).
\end{equation}

 { \subsection{Trap model}
 \label{trap_def}
Many aspects of the glass transition are captured by a family 
of phenomenological models of traps.
We will focus in the simplest case, i. e., the mean-field or 
fully-connected trap model introduced in~\cite{Trap,trap_epl_96}. In this model,
the system is represented by a set of energy wells of 
	  depth $E$ ($E>0$).  The probability density function of wells is 
	  $\rho(E)=(1/T_g) e^{-E/T_g}$. At a temperature $T$, the  system may escape from its well of depth  $E$
 at a rate proportional to $e^{-E/T}$, and will fall in another well of depth $E'$ chosen 
 at random according to $\rho(E')$. It has been shown that, in this case,
 a dynamical phase transition occurs at a temperature $T=T_g$, between a 
 high-temperature ``liquid'' phase and a low-temperature ``aging'' phase~\cite{trap_epl_96}.
For this fully-connected trap model, we define a {\it jump} as the escape
from a trap of depth greater than a threshold $E_t$. Later we compare the properties of these jumps with those corresponding to the 3D-EA model.
We define $\nu_T$ as the number of jumps per unit time; 
$l_T$ as the average energy in a well of depth greater than $E_t$.
$d_T$ as the average time spent in a well with depth greater than $E_t$; and
  $r_T$ as the time interval among jumps.} 
 
 \section{ Macroscopic quantities}
\label{ResMacro}

 \begin{figure}[!ht]
 \centering
   \includegraphics[scale=1.05, clip]{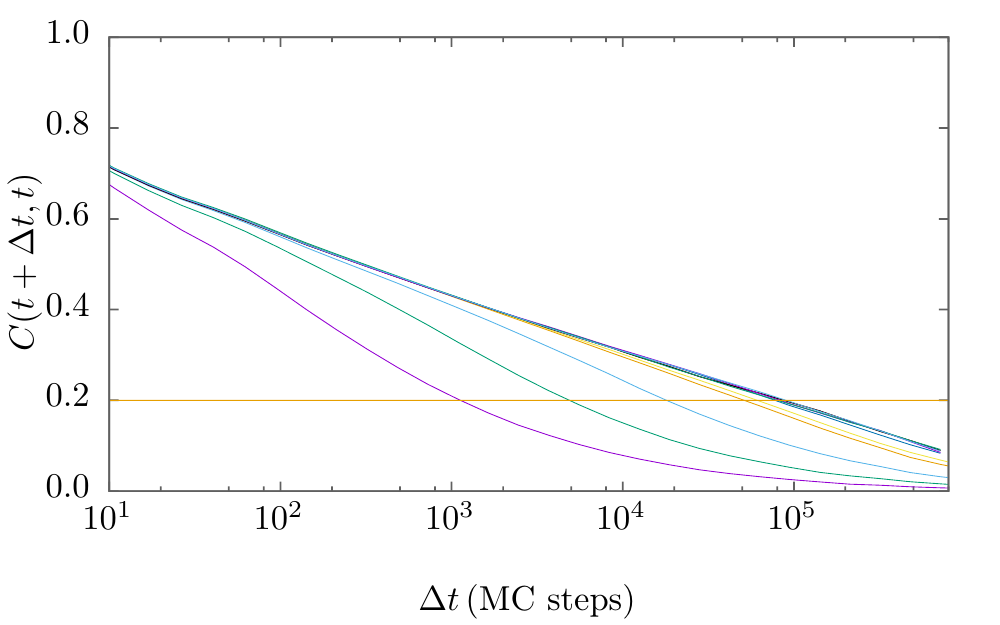}\\
	 \hspace*{-1.7cm} \includegraphics[scale=1.22, clip]{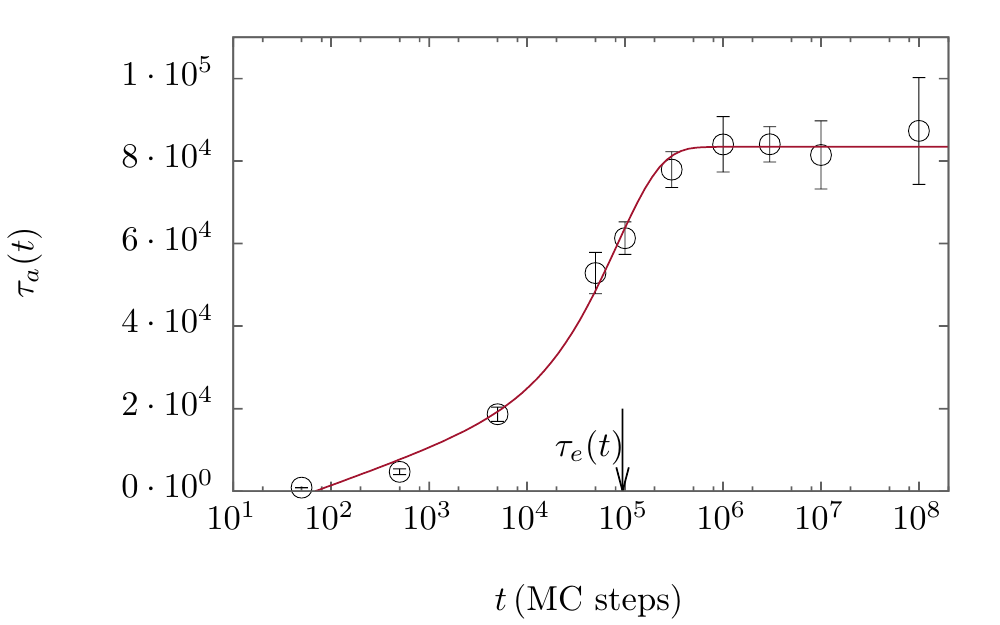}
\caption{(Color online) Top: Global two-time correlations for  {{$T_F=1.2$}}, and 
 several waiting times.  \mbox{$t=5., 5. 10, 5. 10^2, 5. 10^3, 5. 10^4, 10^5, 3. 10^5, 10^6, 
	 3. 10^6, 10^7, 10^8$} MC steps, from left to right. The horizontal line 
 corresponds to $C(t+\Delta t,t)=0.2$. Bottom: $\tau_a$ as a function of $t$ for $T_F=1.2$ 
 (black circles). The (red) line corresponds to the fitting function 
$(C_1+C_2\log(t))e^{-t/C_3}+C_4(1-e^{-t/C_3})$. The arrow indicates $\tau_e=C_3$. }
 \label{Correl}
 \end{figure}

Our first goal is to measure the time it takes to equilibrate the system, using a macroscopic variable. 
{Similar to \cite{VMMLargo}, we find that the energy per spin, $e(t {{,T}})$ is well fitted by $e_\infty + A(T) t^{-b(T)}$ ($e(t,T)$ and this fit
are shown in Appendix), from which a
relaxation time is hard to find, so we decide to use two-times
correlation functions.}

It would be desirable to find $\tau_e^{ideal}$ 
 such  that for $t>\tau_e^{ideal}$,
 $ C(t+\Delta t,t)$ does not depend on $t$.
 We cannot do that due to limited precision in our data.  
 We define  an auxiliary variable $\tau_a(t) $ such that $C(t+\tau_a(t),t)=K$.
 In our case, we choose $K=0.2$. 
 {Notice that an analogous procedure was also employed in \cite{Aging2Eq,Microscopicsio2}}.

 In figure \ref{Correl}-top, we show the two-time correlation 
 as a function of $\Delta t$ for various values of $t$ at $T_F=1.2$.
 In figure \ref{Correl}-bottom, we plot $\tau_a(t)$ for $T_F=1.2$. For $T_F>T_g$, the  relaxation time is then
 $\tau_a^{EQ}(T_F)=  \lim_{t\to\infty} \tau_a(t)$. 
  {Specific  details on how we measure $\tau_a(t)$ are show in 
  Appendix. The main point
 in this section is to get an estimate of macroscopic relaxation time.}

 From figure \ref{Correl}-bottom, we see that  $\tau_a(t)$ grows with $t$ until it
 gets a constant value.
{To describe the growth of  $\tau_a(t)$ at short time, and the final
 constant value,}
 we have fitted $\tau_a(t)$ with 
\mbox{ $F(t)=(C_1+C_2 \log(t))\,e^{-t/C_3}+C_4 (1-e^{-t/C_3})$.}

 From this fit, we can define $\tau_e$, the time at which $\tau_a(t)$ changes from growing to constant,  as $\tau_e=C_3$. Using this procedure, we were able
 to get $\tau_e$ for $T_F=1.5,1.35$ and $T_F=1.2$. 
 Although $T_F=1.0>T_g$ we were unable to estimate $\tau_e$ for 
 this temperature, {since we do not have enough data for long $t$}.
 
 We have checked, for $T_F\geq 1.2$, that $\tau_e$ is a
 good estimate of the macroscopic equilibration time for the one-time 
 macroscopic variables, like total energy and number of flips.

  \section{Mesoscopic and microscopic quantities}
\label{ResMicro}
\subsection{General characteristics of jumps}

{
	Jumps capture, at the box-size scale, the properties of
	the infrequent abrupt changes in energy, magnetization, 
	and decorrelation (as described by overlap decay), which 
	should play a key role in the relaxation process. Indeed, 
	the dynamics become slower with time
	because the system gets stuck in regions of which it is 
	increasingly hard to leave via uncorrelated spin flips. 
	These difficulties can only be overcome with the help of
	cooperative movements that make up a jump; which involve
	only a very small fraction of time intervals.
For instance: $3\%$ for $\gamma=2$, $0.4\%$ for $\gamma=3$ and $0.06\%$ for 
$\gamma=5$; at $T_F=1$. 
Let us remark that during a jump, not only the overlap change is large.
The average absolute changes in energy and 
magnetization of a box are also greater (about $40\%$ and $45\%$, 
respectively, for $\gamma=2$; $50$ and $60\%$ for $\gamma=3$; 
$65$ and $70\%$ for $\gamma=5$) in a jump than in another movement.}
 
\begin{figure}[ht!]
  \centering
	  \includegraphics[scale=.75,clip]{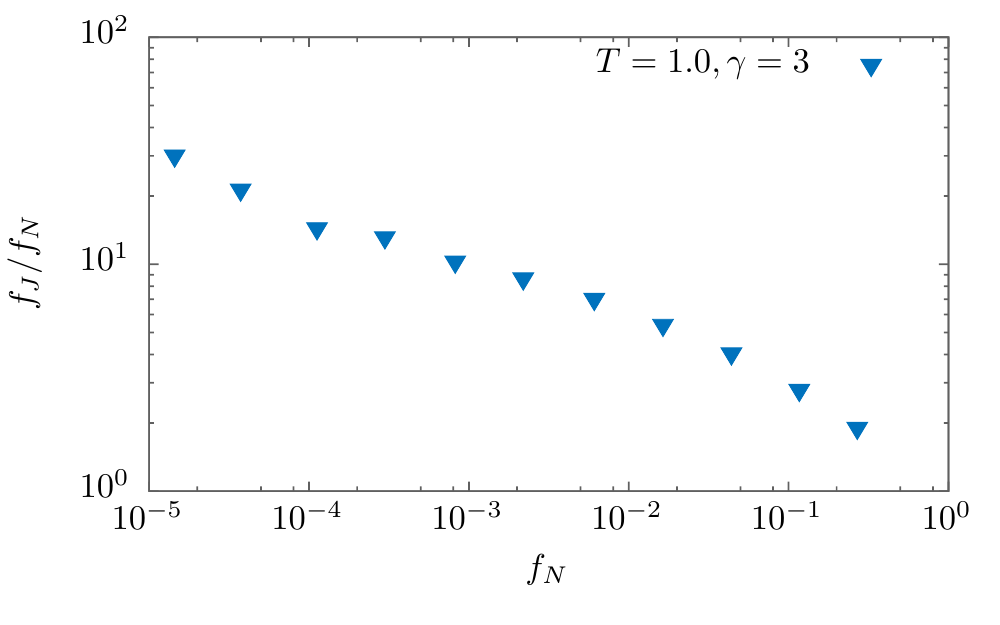}
	    \includegraphics[scale=0.75,clip]{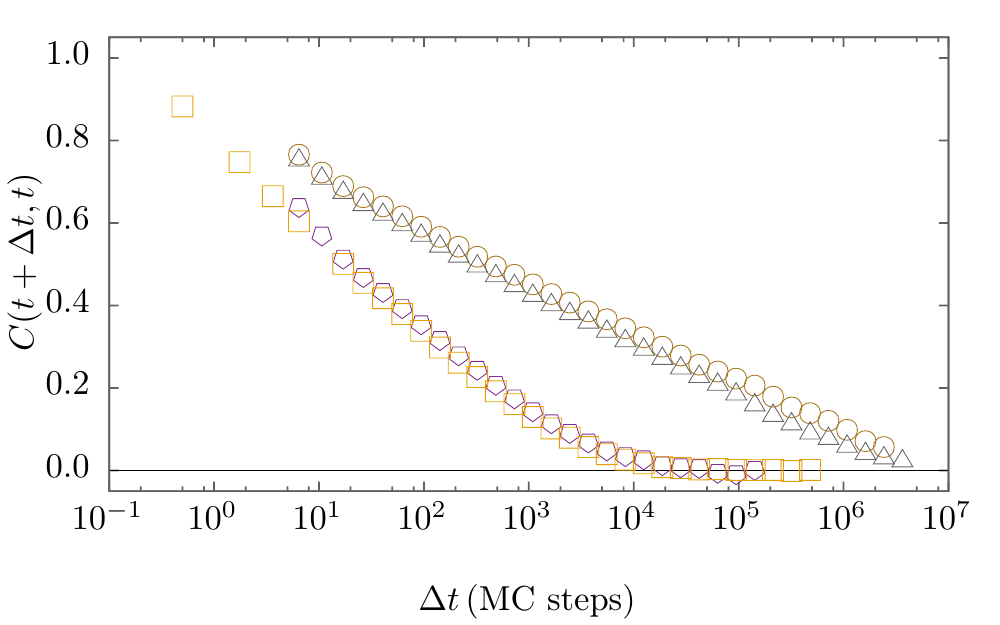}
   \caption{(Color online) Left: The ratio of flip chance on jumps to 
	    flip chance on all time steps, as a 
	    function of the latter, for $T_F=1$, and  $\gamma=3$. Right:  
	    Two-time correlation $C(t+\Delta t, t )$ for $t=10^6$ MC steps. 
	    Diamonds correspond to $T_F=1.5$; triangles to $T_F=1.2$. 
	    Filled symbols show normal relaxation, while empty symbols 
	    show the constrained decay (jumping movements replaced by 
	    normal movements). Here, $\gamma=2$.
	  }
  \label{FC}
  \end{figure}

{
 We expect that a correlation exists between jumps and ''hard to move spins", in the sense that the latter need special coordinated behavior of their 
environments to flip.
To investigate this, 
we have studied individual flips for every spin in
a single box.
On the one hand, we measure the flip chance on all time steps 
($f_N$ ); that is, the number of time steps where a given spin 
has flipped, divided by the whole number of time steps in 
 the simulation. $f_N$ varies from about $10^{-5}$ to $10^{-1}$; i.e., 
 there are spins (the ``fast'' spins) that flip $10^4$ times more than 
 others (the ``slow'' ones). 
 On the other hand, we measure the flip chance within a jump event ($f_J$);
 that is, the analogous to $f_N$ but taking into account only those 
 time steps in which a jump is detected. 
 As expected, $f_N$ is greater for fast spins than for slow ones. 
 Nevertheless, the ratio $f_J/f_N$  has some nontrivial behavior.
 For fast spins, this quantity is about $1.5$, while
 for slow spins, it grows to about $20$.
 In figure~\ref{FC}-left we show the behavior of $f_J/f_N$ as a function of  $f_N$ for
 $T_F=1.0$ and times up to $10^7$ MC steps. Note the strong correlation between 
 jumps and slow spins, reflected in the monotonous decrease of this function.  
 We have obtained similar behaviors for other temperatures; the lower $T_F$, 
 the greater the $f_J/f_N$  ratio.
 }

  { We have also checked that jumps contribute to 
  relaxation more than normal movements, by constraining the dynamics
 in such way that jumps are avoided. We define a constrained dynamics as 
 follows. If after an interval $\Delta t$, a jump is detected, 
 the interval is repeated, i. e., time is reduced in $\Delta t$
 and former configuration is loaded; otherwise the system evolves normally.
 In figure~\ref{FC}-right  we plot $C(t=10^6,t+ \Delta t)$ as a function of $\Delta t$ for
constrained (open symbols) and unconstrained (filled symbols) simulations, 
for $T_F=1.5$ and $T_F=1.2$. We define the relaxation time 
$\tau_{\alpha_C}$  for the constrained dynamics, in an analogous 
way to $\tau_\alpha$.
It is worth mentioning that in the last curve we get that $\tau_{\alpha_C}$ is
roughly twice the value of $\tau_\alpha$. Notice that
while both simulations have run the same number of MC steps, the only
difference is that jumps, which represent just $3\%$
of each boxes flips, are replaced by non-jumping movements in the constrained
model. The ratio  $\tau_{\alpha_C}/\tau_{\alpha}$ increases with $t$, and also 
grows on cooling; meaning that jumps become more relevant both with system 
age and at low temperatures.}

\subsection{Jump statistics}

We analyze how the statistical properties of jump events
evolve with time after a quench.

  \begin{figure}[ht!]
  \centering
   \includegraphics[scale=1.2,clip]{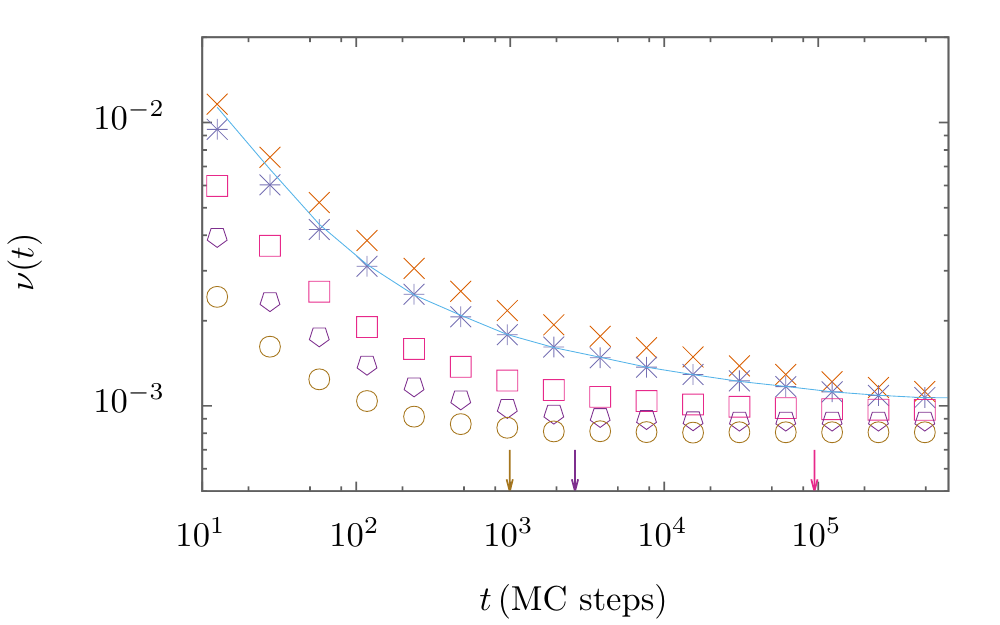}
   \caption{(Color online) Jump frequency as a function of time. Symbols correspond to $T_0=3$, from top to bottom, $T_F=0.9$,
   $1.0$, $1.2$, $1.35$ and $1.5$.
   Line with no symbols is for $T_0=\infty$, $T_F=1.0$.
   Arrows showing
	  $\tau_e$ for $T_F=1.5, 1.35, \mbox{and}\, 1.2$ (from left to right) have been added at figure bottom. Errors are smaller than symbol size.}
   \label{Saltantes}
  \end{figure}

 In figure~\ref{Saltantes}, we show the jump frequency as a function of time 
 in logarithmic scale. For $T_F=1.0$, we have added data for $T_0=\infty$, which
 looks qualitatively similar to the case $T_0=3$. This similarity holds for 
 other values of $T_F$ (not shown).
We see a steep decrease, consistent with the results 
of \cite{Microscopicsio2}; the slope is greater for lower temperatures.
The time for which the decreasing of $\nu$ becomes negligible is compatible with macroscopic relaxation times for temperatures $T_F \ge 1.2$, where we are 
able to measure $\tau_e$. We have indicated these values with arrows at the 
bottom of figure~\ref{Saltantes}, to facilitate comparison.
 
   \begin{figure}[ht!]
   \centering
   \includegraphics[scale=1.2, clip]{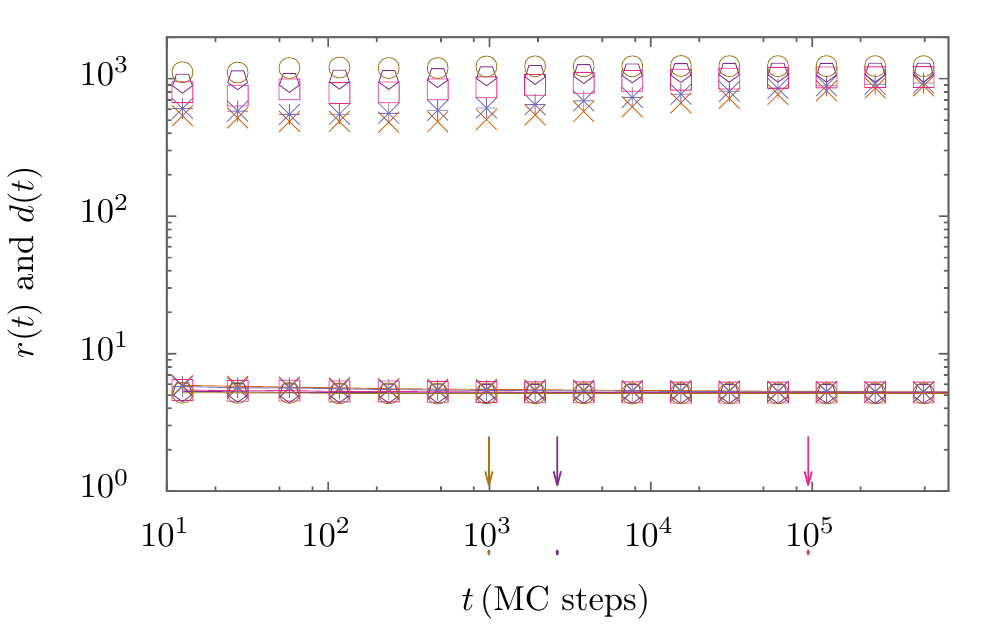}
   \caption{(Color online) Rest time (symbols) and jump duration (lines with symbols) as a function of time.
   For $t>10^5$ there is an artificial decrease of rest time, related to the finite simulation time.
   Symbols and colors are the same as in figure~\ref{Saltantes}. Results 
   for $T_0=\infty$ were avoided for the sake of clarity. Errors are smaller than symbol size.}
   \label{DuracionyEspera}
  \end{figure}

In figure~\ref{DuracionyEspera}, we plot  both  jump duration $d$ and 
   rest time $r$ against time. It is apparent that $d$ is of the order of 
   $\delta t$, which means that most of the jump last one unit time. This 
   is similar to the results in \cite{Microscopicsio2},
where average jump duration is close to time step. Jump duration
decreases very slowly with temperature and has no appreciable time dependence 
even before $\tau_e$.
  However, at odds with 
\cite{Microscopicsio2}, rest time seems to evolve with $t$. This will be 
further discussed in next subsection.
 
  \begin{figure}[ht!]
	  \centering
   \includegraphics[scale=1.2,clip]{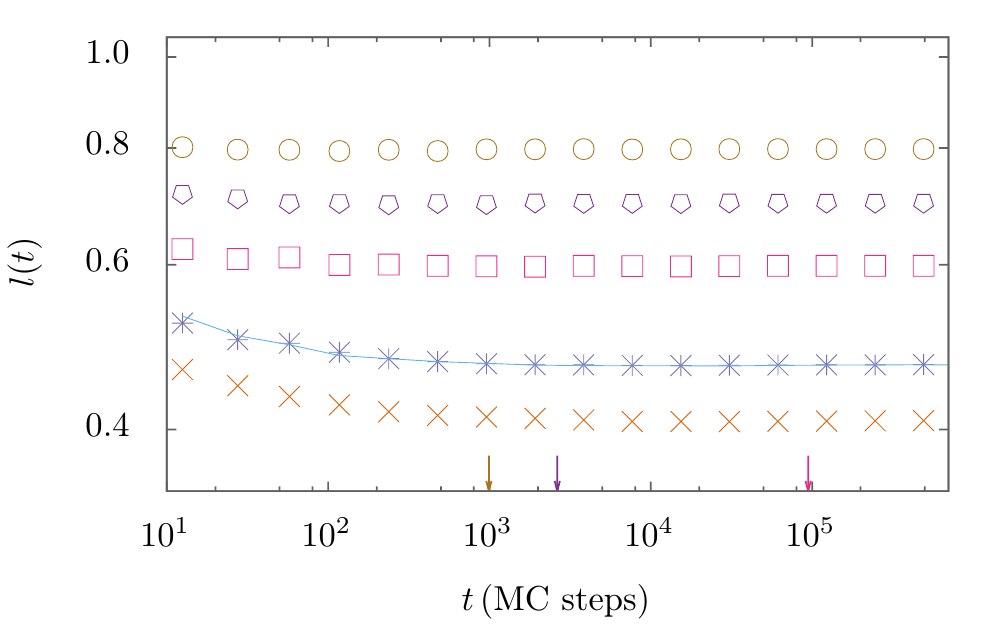}
   \caption{(Color online) Jump size as a function of time for several temperatures. 
	  Symbols and colors are the same as in figure~\ref{Saltantes}. 
	  Errors are smaller than symbol size.}
   \label{DescrSalto1}
  \end{figure}
 
 In figure~\ref{DescrSalto1}, we show how the average value
  jump size $l$ evolves with time. Note that this quantity stabilizes well 
  before equilibration time. Even for $T=0.9$, which is below $T_g$, it does for $t\sim 10^4$.

\begin{figure}[ht!]
   \centering
    \includegraphics[scale=.6, clip]{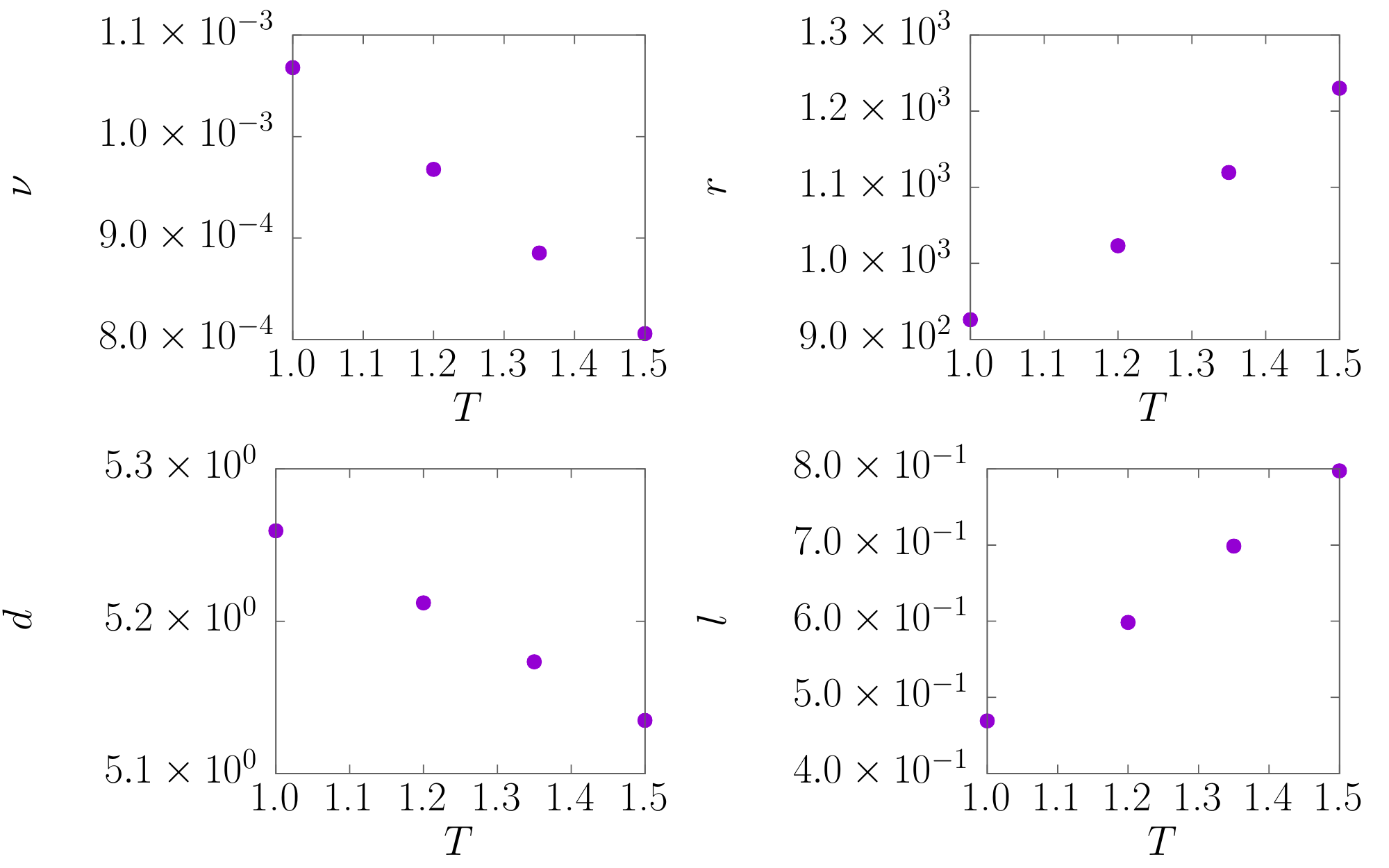}
   \caption{Asymptotic values of $\nu$, $r$, $d$ and $l$ for $T=1$, $T=1.2$, $T=1.35$ and $T=1.5$. Data was taken from results in the range $10^5<t<10^6$ for all temperatures but $T=1.0$, where
   data was calculated in the $10^7<t<10^8$ range. Error bars are smaller than 
	    symbol size.}
   \label{Asintot}
  \end{figure}

{For the sake of completeness, in figure~\ref{Asintot} we show the asymptotic values of $\nu$, $r$, $d$ and $l$ for all the temperatures at which we can equilibrate the system.
 Jump length grows with temperature and jump duration decreases, as might be expected. Jump frequency decreases with temperature and rest time grows,
 which may seem odd, but it can be readily understood if we notice, as
 shown in figure~\ref{Sigma} that $\langle O \rangle$ decreases and $\sigma$ grows with temperature. Then, a jump at higher temperature implies the movement of a grater amount of
 spins.}
  
    \begin{figure}[ht!]
	    \centering
   \includegraphics[scale=1.1,clip]{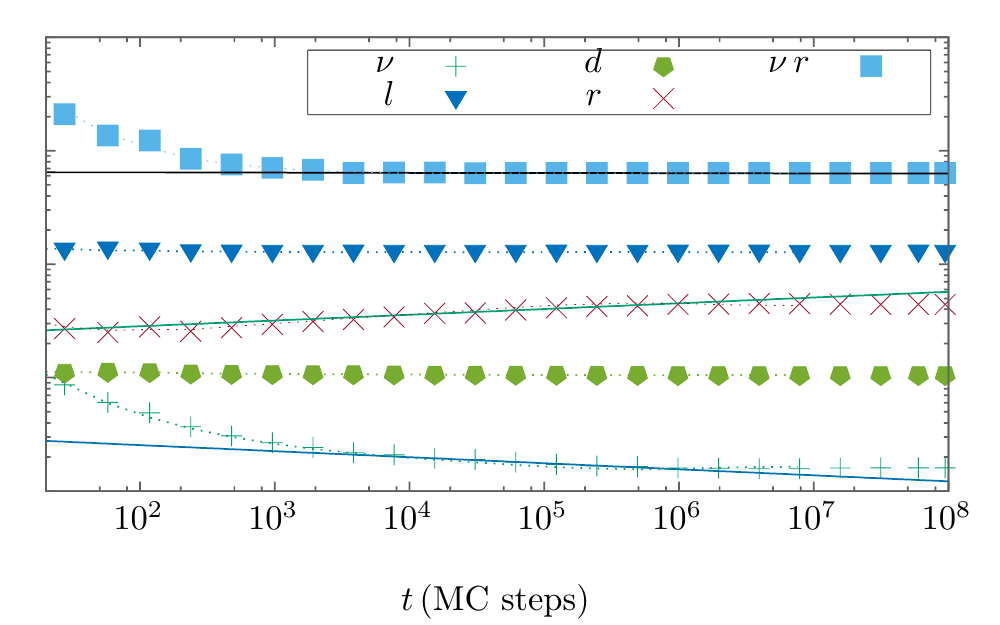}
   \caption{Symbols show results for all observables at $T=1.0$, shifted  
	    so that its comparison becomes more clear. Errors are smaller 
	    than symbol size. Dotted lines show the same results for a bigger 
	    system: $L=32$, $l_b=4$.  Straight lines are the  fit of  
	    $\nu$, $r$, and $r\nu$,  as a function of $t$, using power-law 
	    forms  in the range $10^4<t<5. 10^5$ (for $L=16$ data).  Fitted 
	    slopes for $\nu$ and $r$ are shown in table \ref{slopetable}.}

   \label{Ejemplo1.0}
  \end{figure}

To summarize, at all temperatures, some variables change in a much more 
pronounced way than others. This can be better appreciated in 
figure~\ref{Ejemplo1.0}, where we show all the previous data for $T_F=1.0$ 
in logarithmic scale. Results are shifted (multiplied by a constant) to
facilitate comparison. Let us remark that while the number of jumps decreases 
by about one order of magnitude, other variables have negligible changes 
(as in \cite{Microscopicsio2}),  with the only exception of  rest time, 
which grows at intermediate times. Energy and magnetization jumps (not shown)  
also behave as jump duration or jump length, i.e. are
roughly $t$ independent.

 \begin{table}[ht!]
\centering
 {
 \begin{tabular}{|c|c|c|c|c|}
\hline
\multirow{2}{*}{} & \multicolumn{3}{c|}{$T_F$} \\ \cline{2-4} 
                        & 0.9 & 1.0 & 1.2 \\ \hline
$s_\nu$   & -0.082$\pm$ 0.005 & -0.054$\pm$ 0.004 &-0.013$\pm$ 0.002  \\ \hline
$s_r$  & 0.080 $\pm$ 0.005  & 0.053 $\pm$ 0.004 &     0.013 $\pm$ 0.002 \\ \hline
\end{tabular} }

\caption{Long-time effective exponents for jump frequency and rest time 
	 for $\gamma=3$ (for higher temperature, the average exponents are smaller 
	 than their errors). In all cases, data were fitted in the range $10^4<t<5. 10^5$.}

  \label{slopetable}
\end{table}

\subsection{ Issues with  rest time
 }
 In Ref.~\cite{Microscopicsio2}, the authors study jumping particles, while in this work 
 we study jumps of boxes. These quantities behave similarly, however we have found
 a discrepancy in the behavior of the rest time; which is roughly constant in 
 \cite{Microscopicsio2}. {Similarly, in~\cite{WarrenYRottler1}, it is reported that the first-jump-time probability depends on time, while 
 average persistence times does not. In contrast, in~\cite{Siboni}, it is shown
 that both  quantities depend on the elapsed time since the quench.} Since, according to our simulations (see, for example, 
 figure~\ref{DuracionyEspera}), $r$ is time dependent, some words are in order.

 We will divide our analysis into short, $t< \langle r \rangle$, intermediate, and long times, $t\gg \langle r \rangle$ (and $t> \tau_e$ if $T_F>T_g$). Notice that $\langle r\rangle $ depends on $T_F$ but also on $\gamma$.

For short enough times, i.~e., for $t\ll \langle r\rangle$, most of boxes jump once or never. 
The results corresponding to $r(t)$ in  figure~\ref{DuracionyEspera} indicate that
in this time interval ($\langle t\rangle\simeq 10^3$ MC steps) rest time is nearly 
constant. 
{Nevertheless, the statistics is poor  and biased in this case,
because of the large number of boxes that cannot be considered (as they have 
not even done any jump).}
For longer times ($t\gg \langle r\rangle$), when most of jumping boxes 
do many jumps, it 
should exist some correlation between rest time and jump frequency.
Since jump duration is much smaller than time between jumps, 
the average number of jumps per unit time, multiplied by the rest time,  
should be  equal to the total time multiplied the number of boxes,
i.~e., $r\nu \simeq 1$.
Thus, for long enough times,  we expect that $r$ increases as $\nu$ decreases. 
Notice that this is what happens if figure~\ref{Ejemplo1.0}, where 
{ $\nu$, $r$, $d$,  $l$ but also $\nu r$ have been plotted as a function of $t$
for $T_F=1.0$. After a time of the order of few $\langle r \rangle$, $\nu r$ becomes constant.}

{To get more quantitative, we fit the data with a function $a t^s$, for some constants $a$ and $s$ at 
$t$ about $10^5$ ($10^4<t<5. 10^5$). This is an approximation for
intermediate times (before equilibration, when all slopes become $0$) performed in a limited range of $t$ values.
 } 
 The intermediate-time  ($10^4<t<5\, 10^5$) effective exponents for jump 
 frequency ($\nu(t) \sim t^{s_\nu}$) 
and rest time ($r(t)\sim t^{s_r}$) are opposite ($s_\nu\simeq-s_r$), within 
statistical error, for $T_F=1.0$. The same happens for other temperatures, 
see table~\ref{slopetable}.

For long times, we expect the relation $\nu r=1$ to hold. For $T_F> T_g$, $\nu$ and $r$ will have reached their asymptotic value.
It would be interesting to study whether, for $T_F<T_g$, there is a time when $\nu$ and $r$ stop evolving. Our results
for $T_F=0.9$, up to $10^8$ MC steps, suggest that this is not the case.

 \subsection{Sensitivity of results}
 
In this subsection we present the results of several tests we 
carried out to analyze the robustness of the statistical properties of jump 
events.

We have checked that the outcomes for $T_0=3$ are similar
 to those for $T_0=\infty$. Thus,  most interesting results do not depend 
 much on $T_0$.
  Also, we found qualitatively the same results for 
 different  unit times  $\delta t= 2$, $5$ and $100$,  though choosing 
 smaller $\delta t$ makes the decreasing of $\nu$ more pronounced.

 We have  verified that doubling system size  does not change the results within 
 statistical error: see figure~\ref{Ejemplo1.0}. 
 {
We have measured the coherence length $\xi(t)$ as defined in \cite{WindowManssen}. The
ratio of this length to system size is an important parameter in our study, since
a change of regime, governed by finite size effects, is expected for
  $\xi(t) \simeq L$. In figure~\ref{JL}, we show  $\xi(t)$ for $T_F=1.0$, 
  $T_F=1.2$ and $T_F=1.5$. It is clear that,
   setting $L=16$, we are far from that regime. 
\begin{figure}[ht!]
 \centering
 \includegraphics[scale=1.,clip]{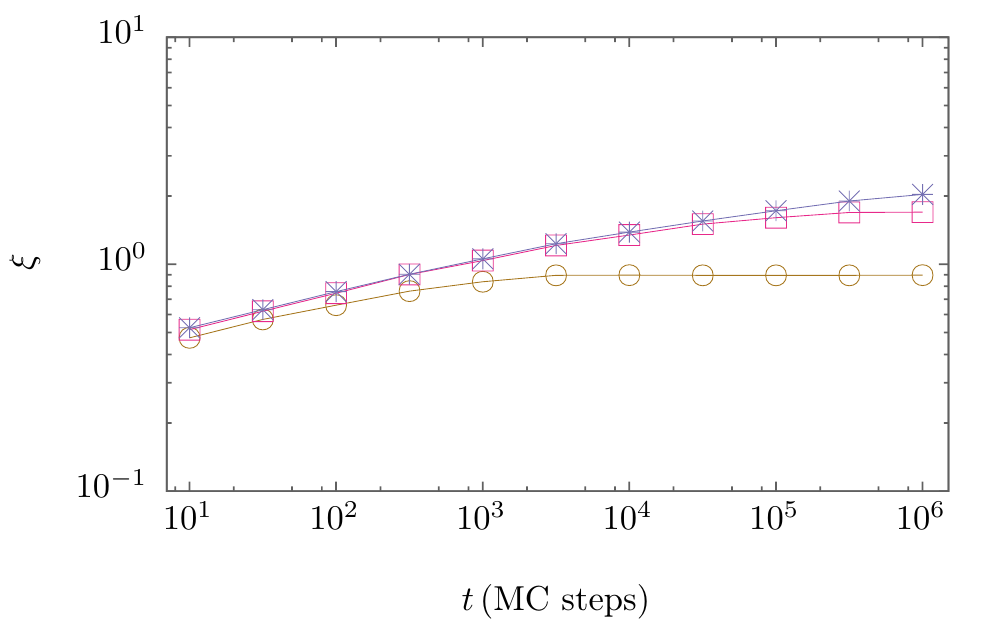}
 \caption{$\xi(t)$ for $T_F=1.0$, $T_F=1.2$ and $T_F=1.5$. Symbol codes are 
	the same as in figure~\ref{Saltantes}. Errors are smaller than 
	symbol size. }
  \label{JL}
\end{figure}
Notice also that while  $\xi$ is a slightly growing function of $t$, $\nu$ is strongly deceasing.
The ratio $l^3/\xi^3$ is always greater than one, meaning that a jump is a pretty infrequent event in which
a representative part of the spins in the box moves cooperatively.
}

\begin{figure}[ht!] 
	   \centering
      \includegraphics[scale=1.1,clip]{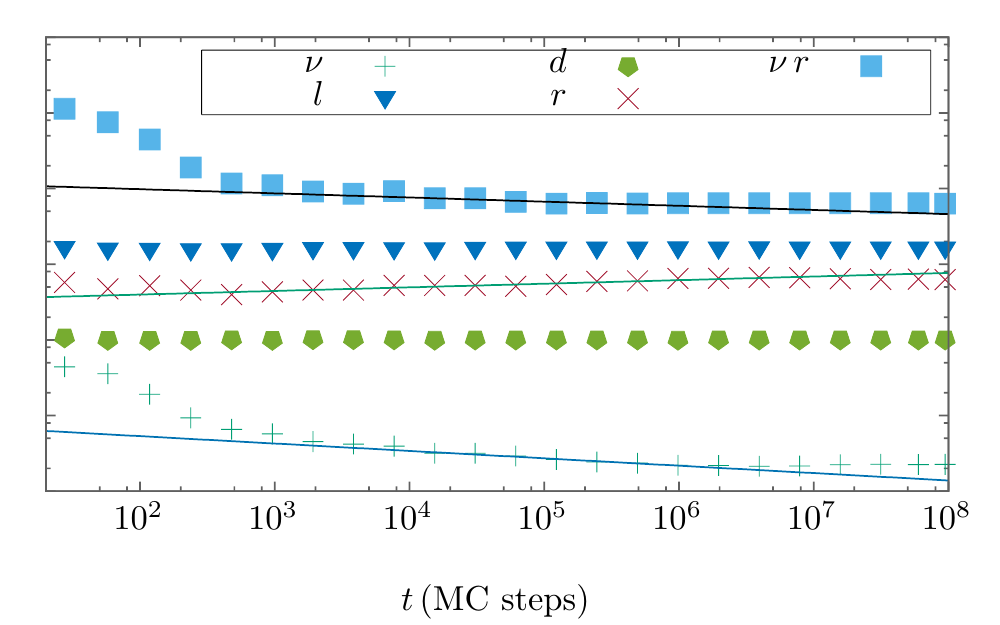}
  \caption{(Color online) Time evolution of jump properties for for $\gamma=5$.
	   Sets of data points are shifted vertically to facilitate comparison. 
	   Let us remark that $\langle r\rangle\simeq 10^5$. Errors are smaller
	   than symbol size.}
  \label{CriteriaCR0} 
\end{figure}

Using an alternative definition of jumps, by changing the value of $\gamma$, also
leads to the same qualitative behavior. For instance, 
a  higher $\gamma$ (more restrictive definition, then fewer jumps) makes 
effective exponent of $\nu$ more negative. For example, we
can compare the results for $\gamma=5$, shown in figure~\ref{CriteriaCR0}, 
with those for $\gamma=3$, in figure~\ref{Ejemplo1.0}. Notice that, in 
first case, $\nu$ decreases by two orders of magnitude, while it decreases 
about one order of magnitude for the second. On the other hand, 
we can observe that the relation $\nu r=1$  does not hold about $t=10^5$.
The reason for this behavior is that, for $\gamma=5$ the average of the rest time 
$\langle r\rangle$ is of the order of $10^5$ (while $\langle r\rangle \simeq 10^3$ for
$\gamma=3$). Now, if we calculate the slope in the range $10^5<t<10^6$, or 
greater, we recover opposite slopes, see table \ref{slopetable2}.

\begin{table}[ht!]
\centering
 {
\label{my-label}
\begin{tabular}{|l|l|l|ll}
\cline{1-3}
\multirow{2}{*}{} & \multicolumn{2}{l|}{\centering Time range} &  &  \\ \cline{2-3}
                  &       $10^4-5 \,10^5$     &       $10^5-10^6$    &  &  \\ \cline{1-3}
    $s_\nu$      &     -0.097 $\pm$  0.011    &      -0.081 $\pm$ 0.007     &  &  \\ \cline{1-3}
   $s_r$    &        0.047 $\pm$   0.014 &         0.082 $\pm$ 0.014  &    &  \\ \cline{1-3}
\end{tabular}

  }

\caption{Effective slopes for jump frequency and rest time 
	 for $T_F=1$, $\gamma=5$ and different time windows}

  \label{slopetable2}
\end{table}

We have also studied \emph{jump energy} and \emph{jump magnetization}, defined as the 
 absolute value of the energy and magnetization differences of each box before and 
 after the  jump. These variables do not sensitively evolve with $t$. The
 energy and magnetization of every box also become stationary well before $\tau_e$.
  Varying $\gamma$, 
 we found that less frequent jumps are related to greater changes in magnetization and 
energy. 

We have explored the use of an alternative criteria of jump, 
with the same threshold for every box. That is, by considering that   
box $i$ is jumping at $t$ when $O_i(t)<B$, with the same constant $B$ for all 
boxes. We find that, since every box has a different quenched disorder,
we get \emph{fast} boxes jumping very frequently and \emph{slow} boxes, jumping few times in the simulation time. Thus, this definition gives some unwanted 
results, related to the fact that rest time is severely underestimated 
at long $t$.

 Finally, we have tried other jump criteria by defining a jump when magnetization 
 differences overcome a certain threshold. We have  found  similar results to the ones 
 presented here using this alternative criteria.
 
The results of these tests give support to the idea that, in spite of the 
details of jump definition, we always get a jump frequency which depends strongly 
on time, but jump properties (with the exception of $r$, at intermediate  times) reach stationarity much before the equilibration time.

 \subsection{Jumps for a trap model}
 
\begin{figure}[ht!]
           \centering
      \includegraphics[scale=1.1,clip]{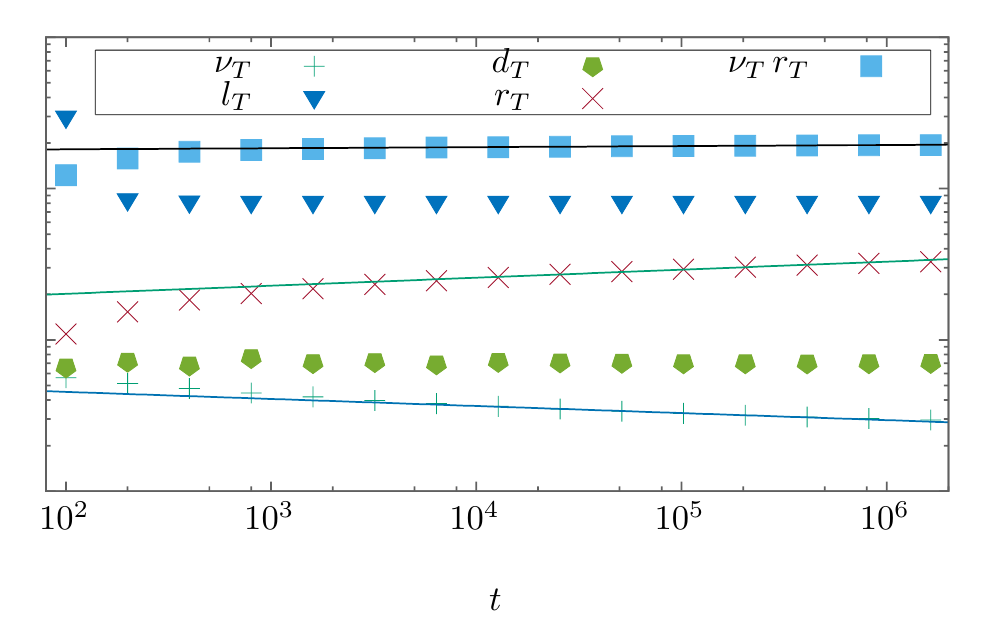}
      \caption{Time evolution of jump properties for the fully-connected
        trap model after a quench from $T=\infty$ to $T_F=1.1\; T_g$. 
	   The jumps are defined using a threshold $E_t=3 T_g$}
  \label{Trap}
      \end{figure}

{ Inspired by the observation that single-particle trajectories near a glass 
 transition are characterized by long periods of localized motions 
 followed by fast jumps, there are attempts to study glass-forming systems 
 in terms of models of non-interacting random walkers. This is the case, for
 instance, of the continuous-time random walk (CTRW) 
 model~\cite{Vollmayr2015,WarrenYRottler1}, based on the assumption that
 jumps are time renewal events, i. e., that the dynamics following
 a jump does not depend on the history. 
 In~\cite{Renewal}, it is argued that CTRW describes activated dynamics, 
 but at high temperatures, or just after the quench, other relaxation mechanisms may be relevant. Interestingly, CTRW model allows for the description of 
 an aging system, even when the movement rules do not depend on time.
 A key ingredient is the implicit synchronization of particles 
 at $t=0$~\cite{Siboni}. 
  
Here we compare the results for the 3D-EA model with those for the 
fully-connected  trap model defined in Sec.~\ref{trap_def}. 
Note the strong similarities between the latter and the CRTW model; 
both considering non-interacting particles, following each a history-independent 
jump  dynamics (because of the  infinite dimensionality, in the case of 
trap model).

  In figure~\ref{Trap} we show the time evolution of the averages of  jump 
  frequency, jump size, jump duration, and rest time between jumps for the 
  infinite dimensional trap model after a quench from $T=\infty$ to 
  $T_F=1.1\;T_g$, and using a threshold $E_t=3 T_g$ in the definition of jumps.
  The same qualitative behavior of these quantities and their analogs for the
  3D-EA model, in figure~\ref{Ejemplo1.0}, is apparent. 
  In the case of the trap model, the average depth $l_T$ of a well leading 
  to a jump, and the average time $d_t$ spent in it, are time independent, 
  while rest time $r_T$ and jump frequency $\nu_T$  are inversely proportional
  to each other.}

\begin{figure}[ht!]
           \centering
      \includegraphics[scale=.75,clip]{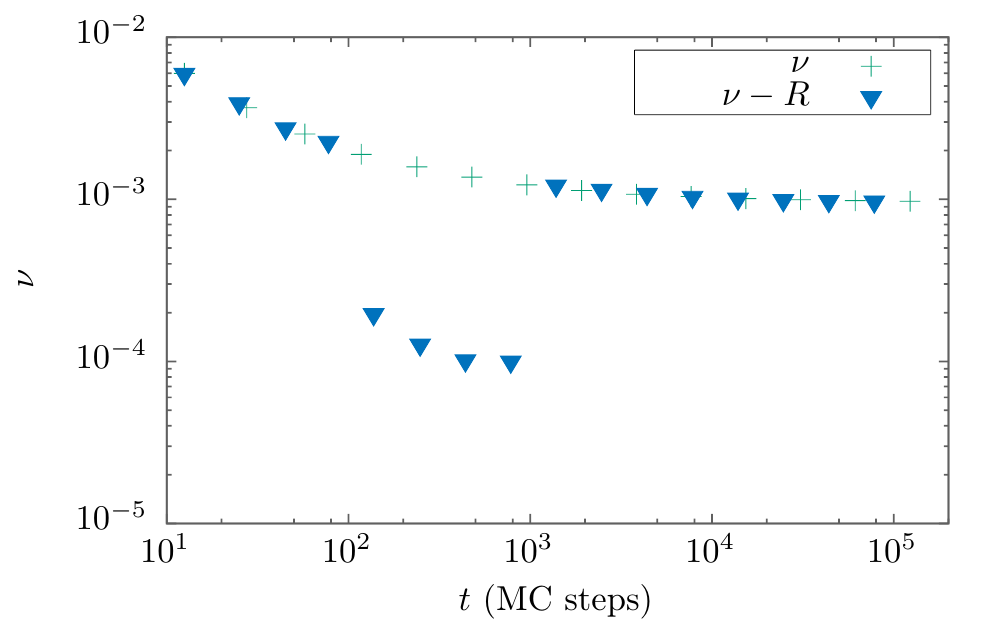}
	\includegraphics[scale=.75,clip]{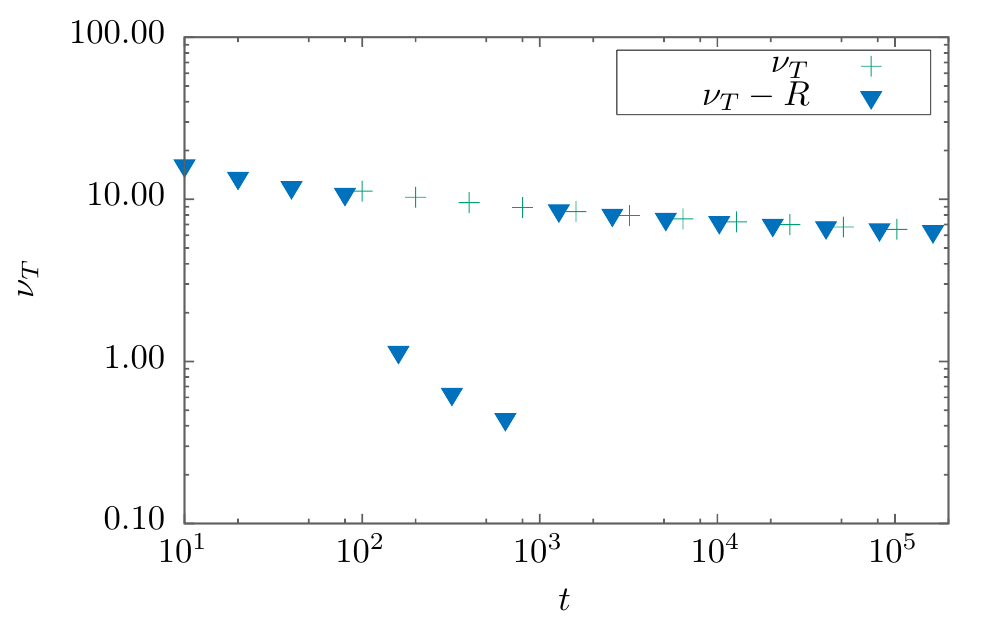}
	\caption{(Color online)
	      Left: Jump frequency (blue triangles) for a 3D-EA model
	      quenched from $T=\infty$ to $T_1=1.2$ at $t=0$ and evolving at this temperatures until $t=100$.  In the interval of time $100<t<1000$ the
	       temperature is lowered to $T_2=1.0$. Then, the temperature is
	       set back to $T_1=1.2$ (for $1000<t<\infty$).
	       The other data points (green pluses) correspond to jump frequency after a single initial quench at $T_1=1.2$.
	       Right: Same as in left figure, for the fully-connected trap model, with $T_1=1.1\;T_g$, $T_2=0.5\;T_g$.}
  \label{rejuv}
      \end{figure}

{In order to test whether the analogy among 
3D-AD and the model of traps can be extended, 
we have performed numerical simulations following a protocol to measure
memory effects~\cite{rejuv_rev}.
Initially, the system is quenched from equilibrium at $T_0=\infty$ to
a temperature $T_1$, at which it evolves for a time $t_1$. Then 
the temperature is suddenly changed to $T_2$, at which the systems evolves
for a time $t_2$. Finally, the temperature is changed back to $T_1$, at which
the system evolves from there onward.
The properties of the system following this protocol, at time 
$t=t_1+t_2+\Delta t$, are equivalent to that of the same system which, 
 after the initially quench evolves always at  $T_1$, for a time
$t=t_1+t_{eq}+\Delta t$. This defines the equivalent time $t_{eq}$,
a measure of the time at temperature $T_1$ for  which the system 
evolves as much as it does for a time $t_2$ at temperature $T_2$.
The equivalent time depends in principle on $T_1, T_2, t_1,$ and $t_2$.
We run simulations for several sets of these parameters, and found
that regarding statistical properties of jumps, it happens that
$t_{eq}\simeq t_2$  for both  EA-3D  and trap models.
This is similar to the result in~\cite{sim_cum_aging}, where rejuvenation 
and memory numerical experiments were study for spin-glass models, and
a cumulative ($t_{eq}>0$) response was found.
As an example, in figure~\ref{rejuv} we show the behavior of jump frequency
following a memory-type protocol (triangles) and after simple quench (pluses),
for both the 3D-EA (left) and the fully-connected trap model (right).
Let us remark the similarity of the behaviors  also in this case.
}

 \section{Discussion } 
\label{Discus}

{ Particle jumps in structural glasses 
have recently received lots of attention. They are useful in relating 
macroscopic observables with microscopic behavior and play an important role
in system relaxation.
In this work, we define jumps for spin-glass models, 
 where the picture of escaping the cage does not apply in an obvious manner. 
Our definition, based on a temperature-dependent threshold for the overlap change between
consecutive configurations, is inspired in the one presented 
in~\cite{Microscopicsio2}. 
Since large overlap changes are related to large 
magnetization and energy changes, alternative definitions of jump using these
latter quantities lead to similar results.
At all studied temperatures, jumps involve a small fraction 
(less than $25\%$) of the spins in the box.
During jumps, at $T_F=1.$, the average chance to 
move ``slow'' spins is about 10 times larger with respect 
to out-of-jumps events (figure~\ref{FC}-left) we observe factors of up to 
100 for the slowest spins at lower temperatures (not shown).
We also show that jumps contribute more than normal movements to overlap 
decay (figure~\ref{FC}-right); their relevance is greater, 
the lower the temperature and the longer the waiting time. Thus, as in the case of structural glasses, 
jump events are closely related to relaxation processes.
}

 { 
 Simulations in the aging-to-equilibrium regime are a useful tool, because they provide glass-like dynamics
in a limited time range, with well known asymptotic values.
 In the dynamics 
of relaxation towards equilibrium, we find that most one-time microscopic observables  
related to jumps become nearly stationary well before $\tau_e$.
One exception is jump frequency, which is a clearly decreasing quantity for times much closer to the equilibration
time. The other exception is rest time. After a few $\langle r\rangle$, $r$ evolves inversely to jump frequency,
and in this sense, it is not an independent quantity. However for $t\ll \langle r\rangle$, $r$ is ill measured and we find a constant value.
Then, our results are qualitatively more similar to those discussed in~\cite{Siboni}
than in~\cite{Microscopicsio2}; the latter reporting a constant rest time for 
a kind of structural glass former. 
The strongly decreasing jump frequency, with almost $t$ insensitive jump lengths and duration, 
first reported in \cite{Microscopicsio2}, is a interesting result. In that work,
 the authors use simulation times of the order of $\langle r\rangle$. It would 
 be desirable to perform molecular-dynamics simulations for that system 
 using longer simulation times, or smaller rest times 
(via less restrictive definition of jumps), and check whether an inverse 
relationship among $r$ and $ \nu$ exists, after a transient of
some $\langle r\rangle$'s.
The asymptotic values of  jump frequency, jump length, jump duration, and rest time, depend on temperature similarly as for structural glass models~\cite{Vollmayr2015}   
}

{The strong similarity between  results in 
figures~\ref{Ejemplo1.0} and~\ref{Trap}, shows that even the simplest 
fully-connected trap model gives an aging-to-equilibrium dynamics in which 
the statistical properties of jumps are qualitatively the same as 
obtained with more detailed models. In this sense, our study gives support 
to the ideas underlying trap models in high dimensional spaces.
By definition, the infinite dimensional trap model neglects 
correlations among consecutive occupied wells, and in this context,
the statistical properties of jumps can be understood as follows.
 In equilibrium at temperature $T$ (long time average), the probability that a well of depth between $E$ and
 $E+\delta E$ is proportional to $\rho(E) e^{E/T} \delta E$.
 Then, at $t=0$, when the system is quenched to a temperature $T_F$ from the 
 equilibrium state corresponding to $T=\infty$, the wells of depth 
 between $E$ and $E+\delta E$  are occupied with a probability 
 proportional to  $\rho(E)\delta E$. This means that, initially, 
 there will be an excess  of occupied wells of low depth, in comparison with
 the  equilibrium distribution at the working temperature.
 As time increases, the trap model relaxes towards equilibrium,
 and the high depth wells become more occupied at the expense
 of low depth wells; leading to a decreasing of the jump frequency.
 This is also the cause of the increasing of the rest time between jumps,
 which evolves as the inverse of $\nu_T$. The departure from this 
 relation observed at short times in figure~\ref{Trap} appears, 
 as mentioned above, because of the poor and biased statistics in this regime.
 On the other hand, since there is no correlation between consecutive jumps, 
 the depth of the well chosen after a jump does not depend on time, 
 neither do $d_T$ and $l_T$, which are determined by this depth.
} 

Memory  and rejuvenation numerical experiments also show that the 
behavior of the individual properties of jumps are qualitatively the 
same for the 3D-EA model as for de fully-connected trap model. 
Further research should be done to 
investigate the extent of this similarity, and the relevance of jump 
correlations to the dynamics of a real system.
 
 \section{Conclusion}
 
 \label{Conclus}
In this work, we generalize the concept of jump, introduced in the context of
glass formers~\cite{Microscopicsio2}, to the case of spin glasses. We divide the system 
into boxes, and define a jump as a cooperative spin flip, making the overlap function 
of a box to decay below a certain amount in a small time interval $\delta t$.
Jumps studied this way collaborate to relaxation more than normal movements.
We study the statistical properties of these jumps as a function of the waiting time $t$
after a quench, for a 3D Edwards-Anderson model of spin glass. 

When this system is quenched to a temperature $T_F$ higher than the glass transition 
temperature $T_g\simeq 0.95$, it reaches equilibrium after a characteristic time $\tau_e$, 
which we determine numerically from the stabilization of the global two-time correlation 
function $C(t+\Delta t, t)$. We confirm that every statistical property of 
jumps becomes independent of time for $t>\tau_e$. At shorter times, some characteristics of jumps have a sensitive dependence on time while others do not.
Jump frequency and rest time may vary on a factor of ten with negligible 
changes in jump length and jump duration, which reach stationarity well before
$\tau_e$.

For  $T_F<T_g$, when the equilibration time diverges, we observe that all the measured 
microscopic observables depend on time, in the time interval that corresponds to our 
simulations. However, while jump duration and jump size change very slowly, jump
frequency decreases much faster. This suggests that in the glass phase the  
number of jumps always decreases but jumps themselves are not $t$ sensitive 
after a relatively short time.

These conclusions do not depend qualitatively on the chosen values of 
$\delta t$, system size, box size nor the details of the criteria to define a 
jump. In particular, if jumps are defined as a function of magnetization or 
energy changes instead of overlap changes, similar results are found.

The statistical properties of jumps  for
the 3D-EA model have qualitatively the same behavior as for the 
fully-connected trap model. This similarity holds also at the level of
rejuvenation and memory experiments. It would be interesting to go 
further in this comparison by exploring correlations among jumps 
for the EA model, and their role in system relaxation.

\section*{Acknowledgments}
We thank UnCaFiQT (SNCAD) for
computational resources.
Data on graphs were averaged using gs\_gav program from glsim package~\cite{glsim}.
This research was supported in part by the Consejo Nacional de Investigaciones 
Cient\'{\i}ficas y  T\'ecnicas (CONICET), and the 
Universidad Nacional de Mar del Plata. JLI is grateful for the financial 
support and hospitality of the Abdus Salam International Centre for Theoretical 
Physics (ICTP), where part of this article was written.

\section*{Appendix: Macroscopic Relaxation Time}

From $C(t+\Delta t,t)$ data, we need to estimate the value of $\tau_a(t)$, the value of
$\Delta t$ for which $C=0.2$. Since data
is subject to statistical error,
it would be desirable to have an analytic expression to fit $C(t+\Delta t,t)$, such that,
for a given $t$, all available information contributes to the fit.
In \cite{VMMLargo}, the shape of   $C_\infty(\Delta t) \doteq \lim_{t\to \infty} C(t+\Delta t,t)$ was
studied for the $\pm J$ Edwards-Anderson Model below $T_g$. Two alternative expressions where proposed: a power law,  $C_\infty(\Delta t)= q_{EA} + A t^\beta$
and a logarithmic $C_\infty(\Delta t)= q_{EA} + {A \over B +\log(t)}$, which performed similarly for their data.
We have tried them setting $q_{EA}=0$. None of them seemed to work for our results in the whole range of times. So we have tried the simplest 
available approximation, $C=A+b \log(t)$.  Clearly, the asymptotic form of this function is nonphysical, so we use it as a
mere approximation of our data on a limited range.  We have restricted our data
to points where $0.1<C<0.4$ for large $t$, and smaller ranges for shorter times.
This approximation gave better results (based on the comparison of goodness of fit divided by the degrees of freedom,
 $\chi^2/n$, where $n$ is the number of points used for the fit). 
We have also checked that  $\tau_e(T)$ do not depend sensitively on the range of the value of $K$.

From $\tau_a(t)$, we have to extract $\tau_e(T)$. Looking at the form of 
the data, figure~\ref{Correl}, we have proposed a simple way to describe it:
 \mbox{$F(t)=(C_1+C_2 \log(t))\,e^{-t/C_3}+C_4 (1-e^{-t/C_3})$}. 
 For short times, $F(t)$ is described by the simplest growing function 
 we could use, $C_1+C_2 \log(t)$, at very long times, it goes to a 
 constant $C_4$. The change of regime (from growing to constant) is 
 given at $t=C_3$, so we estimate $\tau_e(T)=C_3$. This procedure also works 
 for all investigated temperatures above $1.2$.

\begin{figure}[!hbt]
 \centering
 \includegraphics[scale=1,clip]{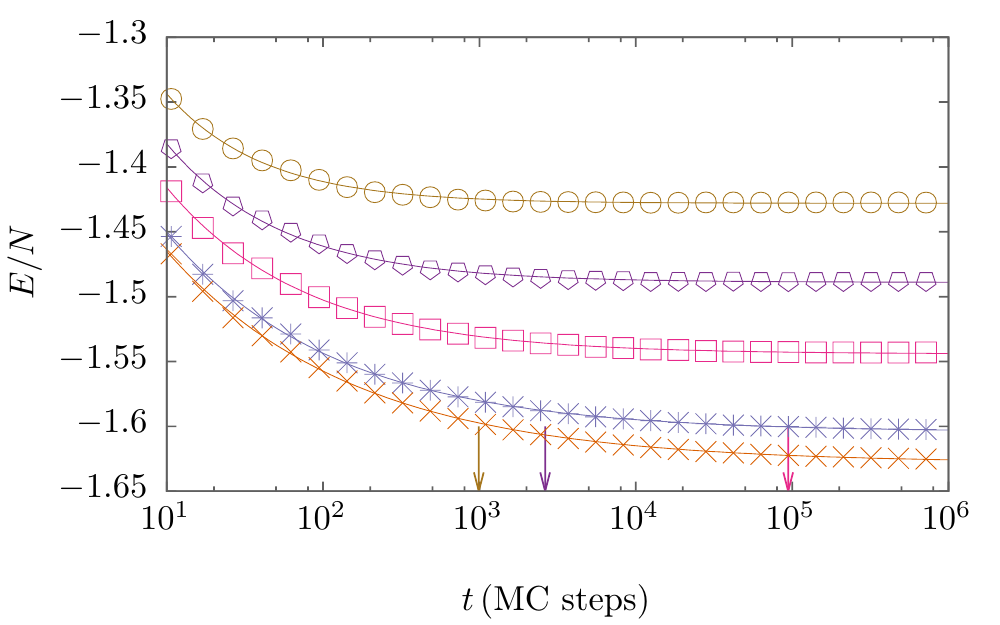}
 \caption{Points: Energy as a function of time. Line: data fit using 
	 $e+A(T) t^b(T)$. Errors are smaller than symbol size.}
 \label{fig_ener}
\end{figure}

As mentioned in the text, $\tau_e(T)$ obtained this way is compatible with energy results, see figure~\ref{fig_ener}.
							 
\section*{References}

\end{document}